\documentclass[utf8]{FrontiersinHarvard} 

\usepackage{url,hyperref,lineno,microtype,subcaption}
\usepackage[onehalfspacing]{setspace}
\usepackage{ulem}

\def\keyFont{\fontsize{8}{11}\helveticabold }
\def\firstAuthorLast{Malakar {et~al.}} 
\def\Authors{Kanaya Malakar\,$^{1}$, Rafael I. Rubenstein\,$^{1}$, Dapeng  Bi\,$^{2}$ and Bulbul Chakraborty\,$^{1,*}$}
\def\Address{$^{1}$Martin A. Fisher School of Physics, Brandeis University, Waltham, MA, USA. \\
$^{2}$Department of Physics, Northeastern University, MA, USA}
\def\corrAuthor{Corresponding Author}

\def\corrEmail{bulbul@brandeis.edu}

\begin{document}
\onecolumn
\firstpage{1}

\title {A Mechanistic Model of the Organization of Cell Shapes in Epithelial Tissues} 

\author[\firstAuthorLast ]{\Authors} 
\address{} 
\correspondance{} 

\extraAuth{}

\maketitle

\begin{abstract}



The organization of cells within tissues plays a vital role in various biological processes, including development and morphogenesis. As a result, understanding how cells self-organize in tissues has been an active area of research.
In our study, we explore a mechanistic model of cellular organization that represents cells as force dipoles that interact with each other via the tissue, which we model as an elastic medium. By conducting numerical simulations using this model, we are able to observe organizational features that are consistent with those obtained from vertex model simulations.
This approach provides valuable insights into the underlying mechanisms that govern cellular organization within tissues, which can help us better understand the processes involved in development and disease.

\tiny
 \keyFont{ \section{Keywords:} tissue, elasticity, organization,  Monte Carlo} 
\end{abstract}

\section{Introduction}

Organ surfaces are covered with confluent monolayers of epithelial cells or endothelial cells, providing physical barriers for organs and bodies. 
Under healthy conditions, the cells in these layers remain static and non-migratory, meaning that they do not move or change position. This feature is an important aspect of their function as a physical barrier, as any gaps or spaces between the cells could allow harmful substances to pass through. Instead, the cells remain securely connected to each other through a network of tight junctions, adherens junctions, and desmosomes, creating a solid-like structure.
The cellular collective behaves as a solid-like object because of the cohesive forces that hold the cells together.  Together, these forces create a strong and stable structure that resists deformation and maintains the integrity of the organ surface.

Inside a confluent tissue, cells form a polygonal tiling of space without any gaps between them. In most cases, they form an amorphous tiling without any spatial order. In spite of major differences between their microscopic constituents and interactions, these cellular assemblies bear a strong resemblance to jammed granular solids. Cells within dense tissues can be considered as particles in a jammed granular solid, where the packing and arrangement of cells play a crucial role in determining the overall tissue properties. This packing can be influenced by various factors, such as cell size, shape, and cell-cell interactions, as well as the mechanical microenvironment.

The essential features that are shared by these two distinct classes of amorphous solids include (i) both form via out-of-equilibrium, non-thermal processes, (ii) their elasticity develops in response to external stresses, and (iii) the cells within tissues are instantaneously in a state of mechanical equilibrium with each cell maintaining force and torque balance.   
These properties define jammed solids: solids whose rigidity emerge in response to external stresses~\cite{Cates_1998a}. A further consequence is that cells in tissues can be represented by force dipoles since force-balance eliminates the force-monopole contribution.
Recent work has uncovered a universal statistical distribution that governs cell shapes in cell monolayers across a wide range of tissues and biological processes~\cite{atia_geometric_2018}. This distribution has been observed in various contexts, including the maturation of the pseudostratified bronchial epithelial layer in both asthmatic and non-asthmatic donors, as well as in the formation of the ventral furrow in the Drosophila embryo. These findings imply a relationship between jamming and geometry that extends beyond specific system details, encompassing both living organisms and non-living jammed systems.

Crucially, tissues differ from jammed granular solids in their display of a complex pattern of cell attributes, including the shapes of cells.
Our primary hypothesis is that, in a solid-like tissue,  the organization of cell shapes is driven by the constraints of mechanical equilibrium. This mechanistic perspective naturally leads to a model of interacting force dipoles~\cite{schwarz_physics_2013}. Recently, this same mechanistic perspective has been framed as an ``emergent theory'' of elasticity of jammed solids~\cite{Jishnu_PRE,Jishnu_PRL}.  From this viewpoint, the organization of cells in tissues and the elasticity of tissues, emerges from the constraints of force and torque balance, locally, as the collection of cells respond to externally imposed forces, which are the natural ``charges'' of this emergent gauge theory~\cite{Jishnu_PRE,Jishnu_PRL}.  The emergent elasticity theory parallels that of classical elasticity, with two crucial differences: physical displacements or strains do not appear, and the elastic moduli are not material properties but depend on how these non-equilibrium, jammed solids are created.

In this paper, we adopt this perspective, however, in practice, our model reduces to the force-dipole models that have been extensively used to study cell-cell interactions in tissues~\cite{schwarz_physics_2013}, if we assume a set of elastic moduli for a tissue.  In future work, we would like to determine these elastic moduli, phenomenologically, through the measurement of stress-stress correlations~\cite{force_chain_gels,Jishnu_PRE}.



The article is divided into the following sections. In section 2, we define the mechanistic model, and discuss how this is related to other models of cell organizations in tissues and the Vertex model(\cite{farhadifar_influence_2007,bi_motility-driven_2016,li_biological_2018,yan_multicellular_2019,das_controlled_2021}). In section 3, we present the results from numerical simulations, and discuss relationships with the vertex model, focusing on the appearance of orientational order and the organization of the magnitudes of cell polarizations (aspect ratios).  In section 4, we present a summary and discussion of future work.



\section{Theoretical Model}

Each cell is modeled as a force dipole, with two equal and opposite forces acting along the same line~\cite{schwarz_physics_2013}. These force dipoles interact with each other by virtue of the fact that they are embedded in an elastic medium, the tissue.   The force dipole associated with a cell is given by the tensor~\cite{bischofs_elastic_2004}:
\begin{equation}
 P_{\alpha \beta} = P \hat{n}_{\alpha}\hat{n}_{\beta} \equiv P q_{\alpha \beta} ~,
 \label{eq:forcedipole}
\end{equation} 
where the cell is viewed as a pair of juxtaposed forces with dipole strength $P$ and orientation $\hat{n}$. 
The dipole strength is a measure of the anisotropy of the cell shape, and can be measured via a cell's aspect ratio~\cite{atia_geometric_2018}. Cell aspect ratio is important because it can influence various cellular processes such as cell division~\cite{gibson2006emergence}, migration~\cite{atia_geometric_2018,mitchel_primary_2020}, and differentiation. The aspect ratio is the ratio of the length of the longest axis of a cell to the length of the shortest axis.

\begin{figure}[h!]
\begin{center}
\includegraphics[width=\textwidth]{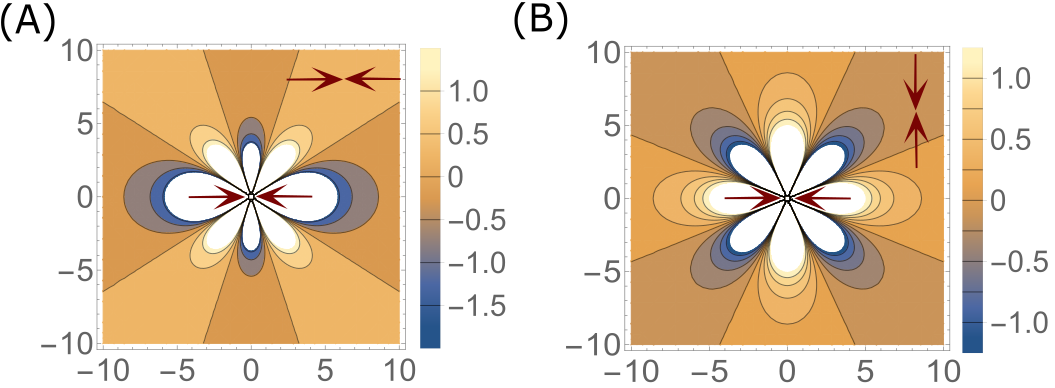}
\end{center}
\caption{ Interaction energy $\varepsilon$ between a pair of dipoles. 
The nature of the energy landscape depends on both the relative distance and the relative orientation of the dipoles. 
In the two figures shown here, one dipole is fixed at the origin with its axis aligned in the x direction. 
A second dipole, either (A) parallel to the first dipole or (B) perpendicular to the first dipole, is placed at different points in space and energy of the configuration is indicated by the color at that point. 
Blue and yellow regions represent attractive and repulsive regions respectively. 
(A) When both dipoles are parallel, the most preferred configuration is the head-to-tail line up. 
Stacking along the y-axis is also a preferred configuration. 
(B) However when the dipoles are mutually perpendicular, both of the previously attractive regions become repulsive. The new attractive regions lie along $45^{\circ}$ to the x axis.}
    \label{fig:greenFunction}
\end{figure}

The elastic energy of the 2D tissue due to interaction between cells modeled here as force dipoles, can be written as $E = \sum_i u_{\alpha \gamma}({\bf r}_i) P_{\alpha \gamma} ({\bf r}_i)$, where $P_{\alpha \gamma}({\bf r}_i)$ is the dipole moment and $u_{\alpha \gamma}({\bf r}_i)$ is the strain field experienced by the $i$-th dipole.  This strain field represents the response to other dipoles (cells) in the tissue, which is modeled as an elastic continuum.
The strain response is, therefore,  expressed in terms of an elastic Green's function
\begin{equation}
u_{\alpha \gamma}({\bf r}_i) = -\partial_{\gamma} \partial_{\delta} G_{\alpha \beta} ({\bf r}_i-{\bf r}_j) P_{\beta \delta}({\bf r}_j ) \equiv G_{\alpha \beta, \gamma \delta} ({\bf r}_i-{\bf r}_j) P_{\beta \delta}({\bf r}_j )
\label{eq:greens_dipole}
\end{equation}
Our analysis is based on the 2D elastic Green's function obtained from the classic Boussinesq solution for an isotropic elastic medium~\cite{zemel_safran_long}:
\begin{equation}
    G_{\alpha \beta}(r) = a_1 \left( a_2 \delta_{\alpha \beta} + \frac{r_{\alpha} r_{\beta}}{r^2} \right) \frac{1}{r}
    \label{eq:greens}
\end{equation}
where the parameters, $a_1$ and $a_2$, are defined in terms of the shear modulus, $\mu$ and the Poisson ratio, $\nu$ as: $a_1=\nu/2 \pi \mu$ and $a_2 = (1-\nu)/\nu$.
The model presented above is based on the assumption of point dipoles, which means that the lengths of the dipoles are considered to be much smaller than the distance between dipoles. Therefore, this purely mechanistic model does not incorporate any steric interactions or geometrical constraints arising from the physical shapes of cells.  
This distinction will be important to bear in mind when comparing the results of this purely mechanistic model to the Vertex model. The explicit form of the dipole Green's function (Eq. \ref{eq:greens_dipole}), obtained from Eq. \ref{eq:greens} is :
\begin{align}
    G_{\alpha \beta,\gamma \delta}(r) = a_1 [ \frac{1}{r^3} (a_2\delta_{\alpha \beta}\delta_{\gamma \delta} - \delta_{\alpha \delta}\delta_{\beta \gamma} - \delta_{\beta \delta}\delta_{\alpha \gamma})
    + \frac{3}{r^5} ( \delta_{\alpha \gamma}r_{\beta}r_{\delta} + \delta_{\beta \gamma}r_{\alpha} r_{\delta}  \nonumber \\ 
    + \delta_{\gamma\delta} r_{\alpha}r_{\beta} + \delta_{\alpha\delta}r_{\beta}r_{\gamma} + \delta_{\beta\delta}r_{\alpha}r_{\gamma} - a_2\delta_{\alpha\beta}r_{\gamma}r_{\delta} ) 
    - \frac{1}{r^7} r_{\alpha} r_{\beta} r_{\gamma} r_{\delta}  ]. \label{eq:dipole_green}
\end{align} 
Thus the elastic energy of the tissue is given by
\begin{equation}
E = \sum_{i<j} P({\bf r}_i) q_{\alpha \gamma} ({\bf r}_i) G_{\alpha \beta, \gamma \delta} ({\bf r}_i-{\bf r}_j) P({\bf r}_j )q_{\beta \delta} ({\bf r}_j)~, \label{eq:interaction}
\end{equation}
where $G_{\alpha \beta, \gamma \delta}({\bf r}_i-{\bf r}_j)$ is the Green's function containing the elastic properties of the tissue.

The results presented in this paper are for an isotropic and incompressible elastic medium, characterized by a Poisson ratio, $\nu=1/2$, and all energies are measured in terms of the  shear modulus, $\mu$. 
It is worth noting that this mechanistic model of cell interactions is inherently tensorial: the interaction strength depends both on the relative positions and on relative orientations of the dipole pair.  In Fig. \ref{fig:greenFunction}, we illustrate this feature, for two representative configurations of force dipoles.


 There is increasing evidence that liquid crystalline, nematic order, and topological defects associated with such order, play a prominent role in cellular development and function~\cite{armengol-collado_epithelia_2022,saw_topological_2017,PhysRevLett.122.048004}.  This liquid crystalline order is associated with the shape of cells, and is viewed as emerging from the collection of cells in tissues behaving as an active liquid crystal. In general, migrating cells tend to have a long axis, and the movement direction of neighboring cells is strongly correlated~\cite{kawaguchi2017topological}. For example, Saw et al~\cite{saw_topological_2017}. demonstrated that defects akin to those observed in nematic liquid crystals manifest in epithelial tissues. The distribution of stress around the defects is comparable to that observed in nematic liquid crystals, and the occurrence of extrusions correlates closely with +1/2 defects. The interaction potential in Eq. \ref{eq:interaction}, derived from a mechanistic perspective {\it of a jammed solid}, has commonalities with liquid-crystal models, and as we show in this work, orientational order can naturally emerge from the interaction between force-dipoles in a jammed solid. This model of force dipoles in a solid has features that are distinct from an active liquid crystal model.  Before embarking on a discussion of our numerical simulations and results, we explore these differences.

 In liquid crystals, the microscopic degrees of freedom are the orientations, $\hat{n}$, of the objects, whether they are ellipsoidal molecules or cells with a longitudinal axis. {Active liquid crystals, are driven out of equilibrium by active stresses and fluid flow~\cite{PhysRevLett.122.048004}}. Nematic order, in both equilibrium and active liquid crystals,  is characterized by the traceless tensor~\cite{Chaikin-Lubensky}:  $Q_{\alpha \beta} =(\hat{n}_{\alpha} \hat{n}_{\beta} -\delta_{\alpha \beta}/2)$ , in 2D.  Rewriting the force-dipole interaction in terms of $Q$, reveals the qualitatively different features of the mechanistic model from the energy function in liquid crystal models (see Supplementary Information for detailed calculation):
\begin{align}
E &= \sum_{i<j} P({\bf r}_i) Q_{\alpha \gamma} ({\bf r}_i) G_{\alpha \beta, \gamma \delta} ({\bf r}_i-{\bf r}_j) P({\bf r}_j )Q_{\beta \delta} ({\bf r}_j)\nonumber \\
& + \sum_i P({\bf r}_i) Q_{\alpha \gamma}({\bf r}_i) h_{\alpha \gamma}({\bf r}_i)~, \label{eq:nematic}
\end{align} 
where we have defined a {\it local field}, $h_{\alpha \gamma}({\bf r}_i)= \sum_{i<j} G_{\alpha \beta, \gamma \beta} ({\bf r}_i-{\bf r}_j) P({\bf r}_j )$.
This representation highlights two important sources of difference between the force dipole model and a generic liquid crystal model: (1) there is an effective field that couples to the nematic field (second set of terms in Eq. \ref{eq:nematic}), and (2) the field $P(\bf{r})$, which is believed to be distributed broadly. 

A universal statistical distribution that governs the shapes of cells in cell monolayers across various tissues and biological processes has recently been discovered in research~\cite{atia_geometric_2018}. This distribution has been observed in different situations, such as the development of the pseudostratified bronchial epithelial layer in both asthmatic and non-asthmatic donors, and the formation of the ventral furrow in the Drosophila embryo. Using the  aspect ratio (defined as the ratio between the long and short axes of the cell) to quantify cell shape elongation, Atia et al~\cite{atia_geometric_2018}  showed that epithelial cells inside confluent tissues obey a universal distribution that is well described by a $k-\Gamma$ distribution~\cite{sadhukhan_origin_2022}. This naturally occuring hetereogenity form the basis for an inherent polydispersity in our model.


An interesting feature, which we will explore in our numerical studies, is the possibility of different dynamics associated with the fields $P$ and $Q_{\alpha \beta}$.  
If the two fields relax on the same time scales, then we have a system with annealed disorder arising from the polydispersity.  
In this case, a generalized nematic order parameter can be defined as  $p_{\alpha \beta}\equiv \langle P Q_{\alpha \beta} \rangle$ .
This is similar to the generalized shape function defined in \cite{huang_shear_2022,fielding2022constitutive,PhysRevE.103.032612,armengol-collado_epithelia_2022}
We refer to this dynamics as Model 1, and unless otherwise stated, the results in the main text are all obtained from numerical simulations of this model.
The configurations {of Model 1 can be fully characterized by the order parameter,  $p_{\alpha \beta}$ }.  
This model is not ``frustrated'' in the classic sense since the interactions are specified just by the elastic Green's function.   
The local field, $h_{\alpha \gamma}({\bf r}_i)$, is strongly influenced by the total magnitude of the polarization, $\sum_i P(\bf{r}_i)$, and as we will show, numerically, the energy of the configurations in Model 1 is primarily governed by the total polarization magnitude of the sample.  
Since this quantity depends on the distribution of $P(\bf{r})$, the energetics also depends on the distribution that characterizes a particular tissue line.  
We will present results, both for the nature of orientational order, and the characteristics of the energy landscape for different realizations of the distribution of $P(\bf{r})$. 

If the field $P$ evolves on a much longer time scale, for example, because the distribution of $P$ is fixed, then we have a situation of quenched disorder.  In this case, the effective interaction is a quenched random variable: $J_{ij}= P({\bf r}_i) G({\bf r}_i-{\bf r}_j) P({\bf r}_j)$.  This model can be frustrated if the effective interaction $J_{ij}$ is incompatible with long range order, which could lead to the appearance of defects in the nematic field.   This scenario is closer to nematic ordering in porous media such as aerogels~\cite{PhysRevE.55.504}, if we threshold $P$ and envision regions with $P$ lower than the threshold as being ``pores''.  In the mechanistic model, the local field, $h_{\alpha \beta}$, is also a quenched variable, and this is a difference from classical nematic models in disordered media.  We refer to this quenched-disorder model as Model 2. 

{In this paper, we focus primarily on the results of Model 1, obtained from Monte Carlo simulations.  We will also present analysis of data obtained from vertex-model simulations of sheared tissues.  We will briefly discuss results from Model 2 in the Supplementary Information.}

\section{Numerical results}

We present numerical results from Monte Carlo simulations of our mechanistic model.
Our primary objective is to understand the interplay between the spatial organization of the magnitude of the force dipoles (cell aspect ratios) and their orientations.
We also explore the energy landscape and reveal important connections with magnitudes of force dipoles.

We compare results from Monte Carlo simulations with those from Vertex model simulations.

\subsection{Monte Carlo Simulation Results}

We perform Monte Carlo simulation to explore the energy landscape and the spatial organization of the force dipoles emerging from the interactions between force dipoles (Eq. \ref{eq:interaction}). 
We consider a square lattice of size $L \times L$ where each lattice site ($r_i$) contains a point force dipole  characterized by a polarization magnitude $P(r_i)$ and an orientation angle $\phi_i$.  
The polarization magnitude is given by the product of the dipole length and the dipole force, $P(r_i)= d F(r_i)$.  In our simulations, the dipole length $d$ is taken to be the same for all dipoles. We also choose $d (\ll l)$, the lattice spacing, to mimic the point dipole in Eq. \ref{eq:interaction}.  The dipole force, $F(r_i)$ is chosen from a distribution, as detailed below.


Several biological cell lines are known to have their aspect ratios distributed according to a $k-\Gamma$ distribution.
We replicate this in our simulations by drawing polarization magnitudes ($P$) from a distribution characterized by parameters $k$ and $\theta$, and described by probability density function 
\begin{equation}
 f(P) = \frac{1}{\Gamma(k) \theta^k} P^{k-1} e^{-P/\theta}.
 \label{eq:kgamma}
\end{equation}
The orientation angles ($\phi$) are randomly chosen between $0$ and $\pi$.
The interaction between force dipoles is given by equation \ref{eq:interaction}, where a dipole's nematic tensor $q_{\alpha \beta}(r_i) = \hat{n}_{\alpha} \hat{n}_{\beta}$ can be obtained by writing the dipole orientation vector as $\hat{n}=(\cos \phi , \sin \phi)$. 
The total energy is obtained by summing the long range interactions between all pairs of dipoles with periodic boundary condition on all sides.\\

To analyze the energy landscape, we access a multitude of metastable states by thermalizing the system at a high temperature and then {running the simulation at a temperature that is low compared to the energy of the system: $\frac{k_BT}{E} \approx 10^{-6} $.}
For convenience, we will often use a dimensionless form of the energy given by $\varepsilon = \frac{El^3 }{N\langle P \rangle^2 d^2}$, where $E$ is the dimensionful energy (Eq. \ref{eq:interaction}), $l$ is the lattice spacing, $N$ is total number of dipoles, $\langle P \rangle$ is the mean value of polarization magnitude, and $d$ is the dipole length.

During each Monte Carlo step, the two properties of each dipole, polarization and orientation, are updated following two different rules. 
For the orientation, we employ model A (non-conserved) dynamics [\cite{hohenberg_theory_1977}], where a new orientation angle is proposed randomly between 0 and $\pi$.
The polarization magnitudes are updated following model B (conserved) dynamics [\cite{hohenberg_theory_1977}], where the polarization magnitudes of two randomly chosen dipoles are exchanged.  This dynamics guarantees that the overall distribution of the polarization does not change during the simulation, however, the spatial organization of the magnitudes evolves along with the orientation of the dipoles.
We draw the polarization magnitudes of our force dipoles from a k- distribution for each of our initial conditions.  Individual Monte Carlo runs then lead to a sampling of the energy landscape corresponding to a particular k- distribution.  Statistical properties can then be inferred by averaging over initial conditions.
Acceptance rate of a proposed update is determined by the Boltzmann factor $\exp(-\frac{\Delta E}{K_B T})$, $\Delta E$ being the difference of energy of the proposed state and the present state, and $T$ is the temperature.\\

The simulation is run for a long enough time to ensure that the system reaches a metastable, local minimum, of the energy function: the energy fluctuates around  a well-defined average value. The time evolution of the system is monitored by observing three different quantities - (i) the total energy of the system, $E$, which  undergoes a sharp decrease from the  initial high-energy state, and then fluctuates around a much lower value,   (ii) the average nematic scalar order parameter, $S =\langle \cos 2\phi \rangle$, which evolves from $\approx 0$ in the initial  state to a value close to unity in the low energy state, (iii) the average weighted order parameter, $\Omega =\langle P_i \cos 2\phi \rangle/\langle P \rangle$, which  shows a similar trend, growing from $\approx 0$ in the initial state to a large positive value ($\geq 1$) in the final state. The time evolution of these three quantities, for a sample run,  are plotted  in fig \ref{fig:MonteCarlo}.\\

\subsection{Orientational Order} 
\begin{figure}[h!]
\begin{center}
\includegraphics[width=0.8\textwidth]{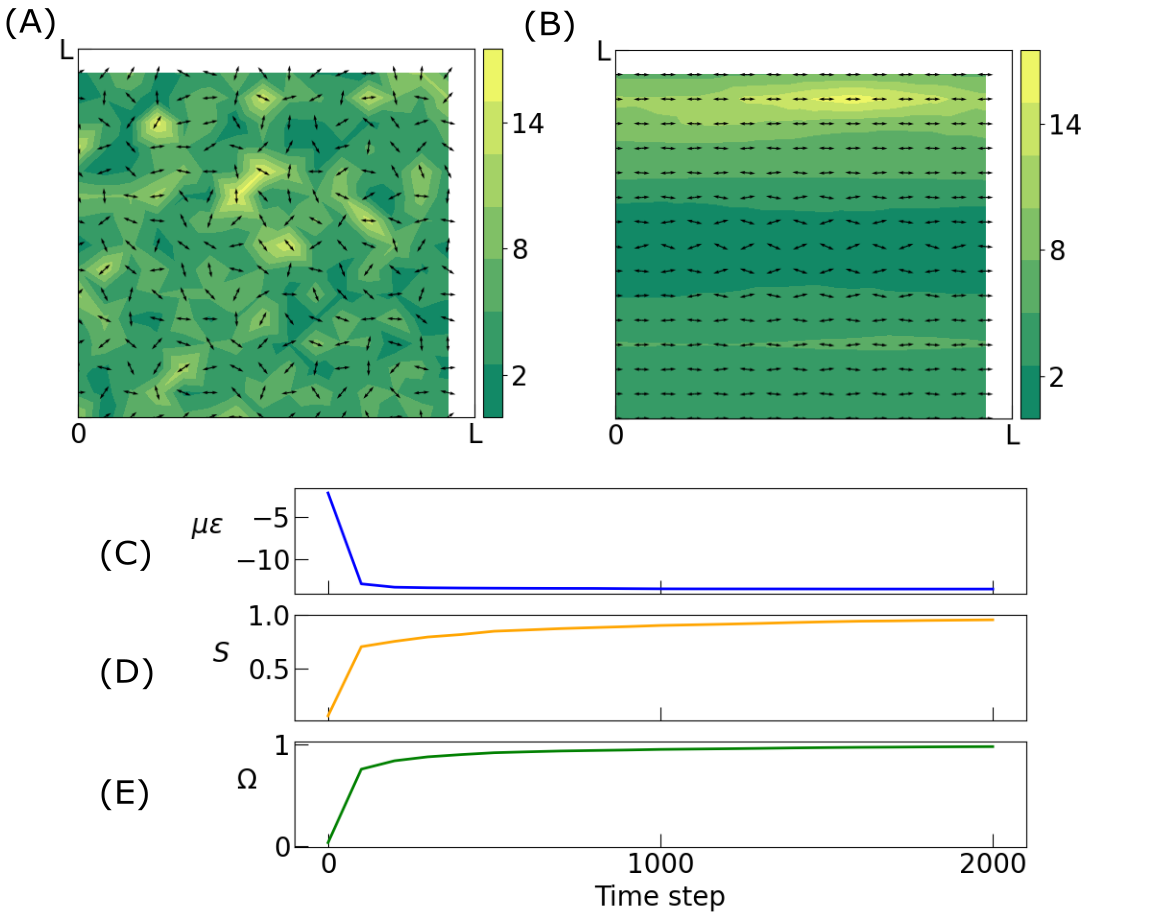}
\end{center}
\caption{ (A) Initial and (B) final stages of a Monte Carlo simulation of dipoles on a square lattice. 
Arrows indicate orientation of dipoles, and colorbar indicates magnitude of polarization. 
Polarization magnitudes are randomly drawn from a gamma distribution ($k$=2.5, $\theta$=2.0). 
(A) In the initial state dipoles have random orientations and there is no spatial correlation between polarization magnitudes of neighboring dipoles. 
(B) The final state obtained by slowly anenealing the system to a low temperature, shows formation of domains of similar polarization magnitudes and domains with non-zero nematic order.
(C) During the anneal, the energy of the system decreases sharply and soon the system gets stuck in a metastable state. 
(D) Average nematic order parameter $S$ shows a steady shift from zero to $\pm1$ indicating emergence of nematic order. 
Here $\phi$ is the orientation angle of each dipole. 
(E) The weighted order parameter $\Omega$ is plotted against time. 
In the disordered state the value of this weighted nematic order parameter is close to zero, but as the system gets ordered its value becomes more positive or negative. 
In this example, this order parameter approaches unity.
Simulation parameters : $k=2.5, \theta=2.0, l=0.1, L=1.5, k_B T=0.001, \nu=0.5, \mu=0.5$.}
    \label{fig:MonteCarlo}
\end{figure}

Figure \ref{fig:MonteCarlo} shows initial and final configurations of a sample run.
The initial state (A) shows dipoles that are placed on the sites of the square lattice, with random orientations and polarization magnitudes drawn from a $k-\Gamma$ distribution.  
The final state (B), which has a much lower energy, has a majority of the dipoles aligned along the horizontal direction. 
We also observe that the dipoles are sorted into well-defined domains by their polarization magnitudes.

\begin{figure}[h!]
\begin{center}
\includegraphics[width=\textwidth]{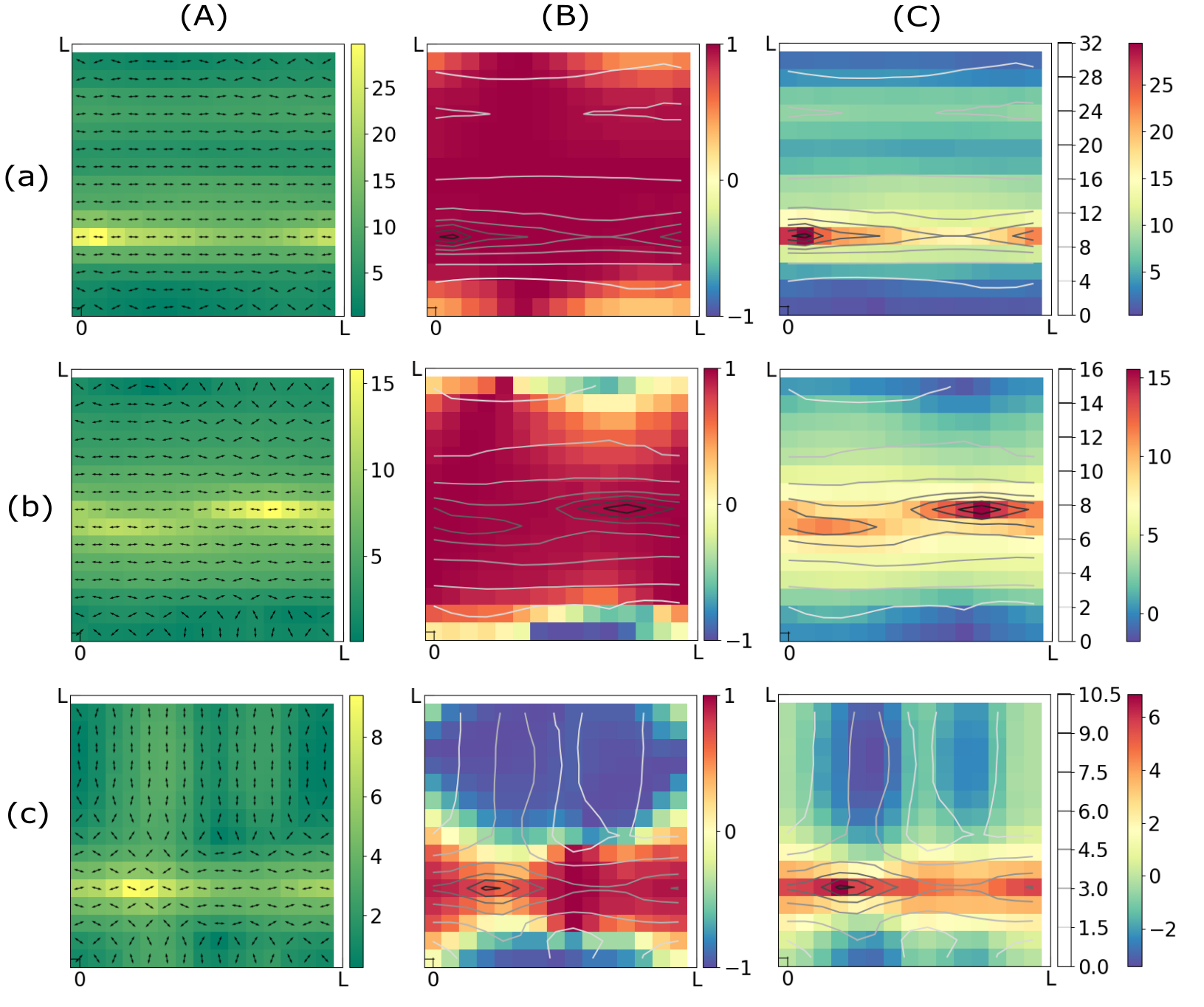}
\end{center}
\caption{ Snapshots of final states obtained from Monte Carlo annealing runs starting with three different initial realizations of the distribution of $P$. 
(a) $k=2.5$, $\theta=3.0$, (b) $k=2.5$, $\theta=2.0$, (c) $k=2.5$, $\theta=1.0$. All three columns (A-C) of each row (a,b,c) shows the same configuration through different lenses. 
Column (A) shows heatmap of polarization magnitudes and orientations of dipoles. 
Column (B) represent the nematic scalar order parameter $S = \cos 2\phi$ as heatmap. 
The heatmap in column (C) shows the weighted nematic order parameter $\Omega = P \cos 2 \phi / \langle P \rangle$. In columns (B) and (C), the polarization magnitudes are represented by contour lines. 
It is evident from column (C) that the domains of large polarizations align with largest absolute values of $\Omega$ and hence are coincident with ordered domains. 
Simulation parameters: $\nu=0.5$, $\mu=0.5$, $L=1.5$, $l=0.1$, $k_B T=0.001$ .}
    \label{fig:ModelAsnapshots}
\end{figure}

In fig \ref{fig:ModelAsnapshots}, we take a closer look at the {\it spatial organization} of  the order parameters $S$ and $\Omega$ in the final, low-energy states. 
The three rows represent results from simulations with three different values of $\theta$ (Eq. \ref{eq:kgamma}). 
Each row represents  adifferent representation of the same end state, in order to highlight different measures of the ordering. 
The leftmost figure in each row shows the two defining properties of every dipole - orientation (arrows) and polarization magnitude (heatmap). 
We see clear formation of domains of large polarization in all three sets of data.
The middle figure in each row shows local values of scalar order parameter $S$ which again organizes the system into domains of nematic order, oriented in different directions.
In the rightmost figure of each row, the heatmap shows local values of the weighted nematic order parameter $\Omega$. 
From the figures it is clear that $\Omega$ provides a better characterization of the ``order'' than the pure nematic order parameter, 
$S$, and clearly identifies regions where the nematically ordered domains overlap with domains of large polarization magnitudes.
For clarity we have also plotted the polarization magnitudes using contour lines.

These figures clearly  demonstrate, one of our primary findings, that domains of largest polarization magnitude overlap with domains of highest nematic order.   
Although not surprising in retrospect, this aspect of the mechanistic model is missing from pure liquid crystalline models of tissues where the magnitude of P is the same for all cells, and points to the need for characterizing cell-polarity in tissues by both the magnitude of the polarization and its orientation.  
Our results imply that because of mechanistic interactions between cells, defects in orientational order would likely be localized to regions where the cells are not significantly distorted from an isotropic shape. 
The consequences of this for tissue development and morphogenesis needs to be explored further. 
Therefore, further exploration of the mechanisms that regulate cell orientation and the consequences of defects in cell orientation for tissue development and disease is crucial. This could involve using advanced imaging and computational techniques to study the dynamics of cell orientation in tissues, as well as investigating the role of mechanical and biochemical factors in regulating cell orientation. Ultimately, a better understanding of these processes could lead to new therapeutic strategies for treating a wide range of diseases and disorders.

In a separate set of simulations, of Model 2, we keep the dipoles fixed at the same positions as in the initial disordered state. Dipoles do not exchange polarization magnitudes and are allowed to update only their orientation angles.  Please see supplemental data for results.

The dynamics of cells within solid-like biological tissues is, of course, much more complex than can be captured by Model 1 or 2, discussed in this work.  
To connect to observations in a much more realistic model of tissues, we present results from simulations of the vertex model, subjected to external shear. Most biological tissues undergoing shape change due to development, growth, morphogenesis, wound healing etc. are under external stresses. We have not extended our mechanistic model to explore the effect of external stresses.  Below, we compare the order-parameter correlations that develop in a vertex model under shear to results from Model 1. 

\subsection{Vertex Model Simulations}

The Voronoi-based implementation~\cite{bi_motility-driven_2016} of the vertex model~\cite{farhadifar_influence_2007,li_mechanical_2019,li_biological_2018,yan_multicellular_2019,mitchel_primary_2020,das_controlled_2020} is employed to model a 2D cell layer, with the cell centers ${\mathbf{r_i}}$ serving as the degrees of freedom and their Voronoi tessellation dictating the cellular structure\cite{bi_motility-driven_2016}. The mechanics of the cell layer are described by the energy function~\cite{staple_mechanics_2010} 
\begin{equation}
E= \sum_{i=1}^N \left[ K_A (A_i-A_0)^2+ K_P (P_i-P_0)^2 \right]
\end{equation}
, where $N$ is the number of cells, $K_A$ and $K_P$ are the area and perimeter elastic moduli, respectively, and $A_i$ and $P_i$ are the area and perimeter of the $i$-th cell. $A_0$ and $P_0$ are their corresponding equilibrium values. The origin of the first term in the expression, which is quadratic in the cell areas ${A_i}$, can be attributed to the incompressibility of the cell volume. As a result, a 2D area elasticity constant $K_A$ and a preferred area $A_0$ are produced, as discussed in ~\cite{farhadifar_influence_2007,staple_mechanics_2010}. The second term in the expression, which is quadratic in the cell perimeters ${P_i}$, is a result of the contractile nature of the cell cortex and is described by an elastic constant $K_P$~\cite{farhadifar_influence_2007}. The target cell perimeter $P_0$, which represents the interfacial tension between adjacent cells arising from the competition between cortical tension and adhesion ~\cite{farhadifar_influence_2007,staple_mechanics_2010}. Here, we focus on the case where all cells share the same single cell parameters: $K_A, K_P, P_0, A_0$. We also choose $A_0 = \bar{A} $, the mean cell area, which defines the unit of length. For all results presented in this work, we used  $N=400$ cells. 

To study tissue mechanical response in the vertex model, we subject the model tissue  to quasistatic simple shear  using Lees-Edwards boundary conditions~\cite{Tildesley_book, huang_shear_2022}.  We start from a strain-free state, the strain $\gamma$  is increased in increments of  $\Delta\gamma=2\times10^{-3}$, while   cells are subject to an affine deformation given by $\Delta \mathbf{r_i} = \Delta\gamma \ y_i \ \hat{x}$.
At each strain step,  the tissue energy is minimized to find a mechanically stable configuration.

\begin{figure}[h!]
\begin{center}
\includegraphics[width=0.8\textwidth]{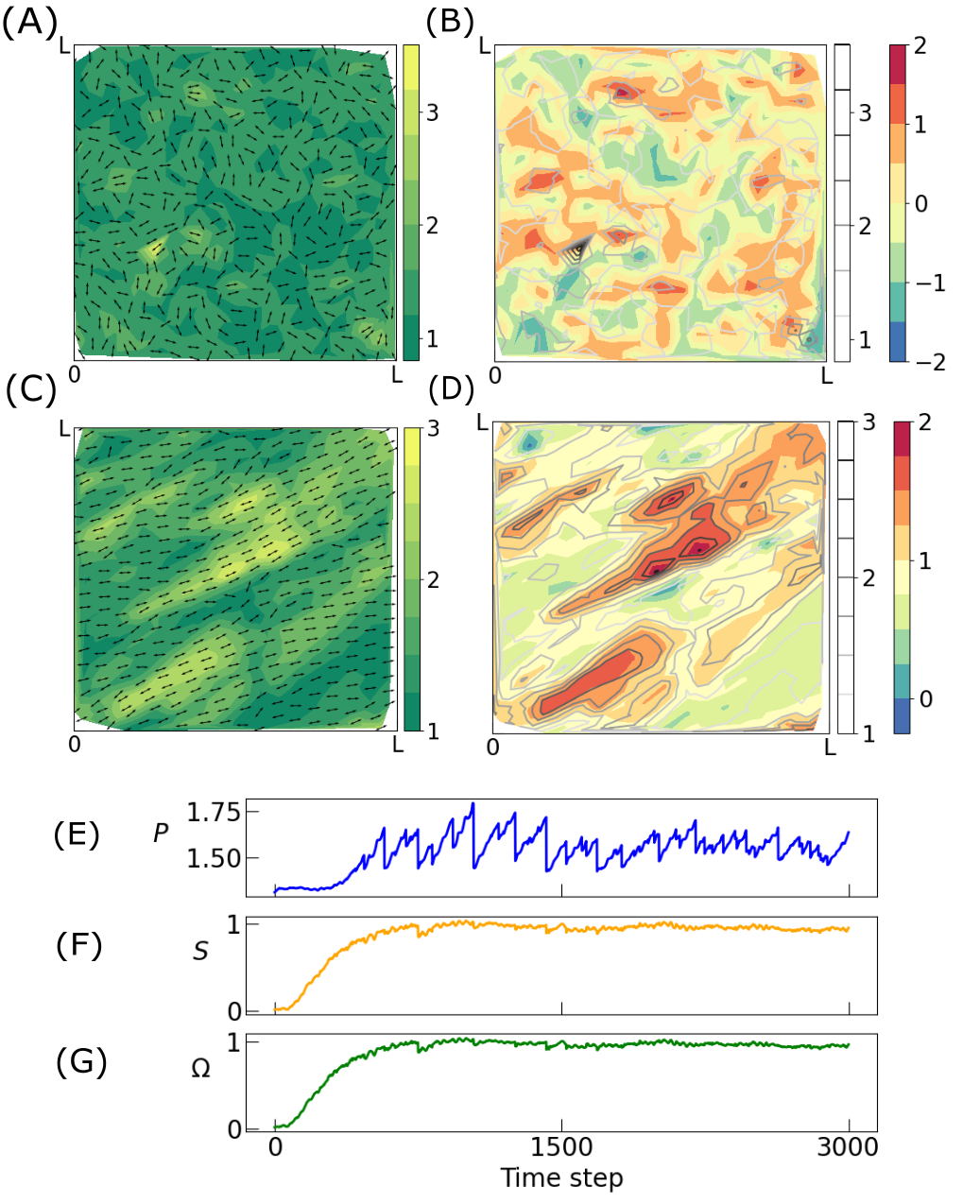}
\end{center}
\caption{Panel showing results from voronoi Vertex model simulations where a constant shear in $xy$ direction is applied to a tissue which undergoes multiple rearrangements in response to the external shear. 
Snapshots from the vertex model simulations from (A, B) initial and (C, D) final time step. 
(A, C) shows orientation (arrows) and magnitude of polarization (heatmap), and (B, D) shows weighted nematic order parameter (heatmap) and polarization magnitude (contours). 
In the initial state, the system is random with no spatial or orientational order, but as time progresses we observe emergence of domains with large nematic order which also coincides with domains of large polarization. 
This matches with our observation from the Monte Carlo simulations above.  
(E) Distribution of polarization changes throughout the process unlike the Monte Carlo simulations where polarization distribution is held constant throughout an annealing run. These fluctuations are characteristic of a steady-state flow at yield stress, which proceeds via a sequence of elastic loading and plastic failure.
Both (F) nematic scalar order parameter $S$ and (G) weighted nematic order parameter $\Omega$ goes from near zero in the random phase to a large positive value in the ordered phase.}
    \label{fig:Max_configuration}
\end{figure}

{
We plot the data obtained from vertex model simulations in fig~ \ref{fig:Max_configuration}. 
Fig~ \ref{fig:Max_configuration}(A) shows initial state ($\gamma = 0$) of the tissue with the cellular aspect ratios indicated by the heatmap and cell orientations indicated by arrows.
Fig~ \ref{fig:Max_configuration}(B) shows a configuration of the tissue in the quasistatic plastic flow regime at larger strain values. Here the cell shapes are oriented  at $\sim~45^{\circ}$ to the $x$ axis due to shear.
We observe alignment of cells and formation of domains of large polarization magnitudes parallel to the direction of the external shear.
Figs~ \ref{fig:Max_configuration}(C,D) shows the corresponding heat maps for the order parameter $\Omega$, and contour lines indicating magnitude of polarization.
Here again we observe that the domains of large polarization coincides with domains of large nematic order, similar to the Monte Carlo simulations.
This system reaches steady state following a very different mechanism from the Monte Carlo simulations. 
In the steady state, the cells undergo elongation and plastic failure~\cite{huang_shear_2022} multiple times as indicated in the time series plot of the mean polarization magnitude.
Still we find end states that are qualitatively similar to those of annealed Monte Carlo runs.
The emergence of order in the system is indicated by the time series plots of $S$ and $\Omega$, both of which starts at a value close to $0$ in the initial state and plateaus to a large positive value as time progresses. 
}

\subsection{Comparison between Monte Carlo and Vertex model}

\begin{figure}[h!]
\begin{center}
\includegraphics[width=0.9\textwidth]{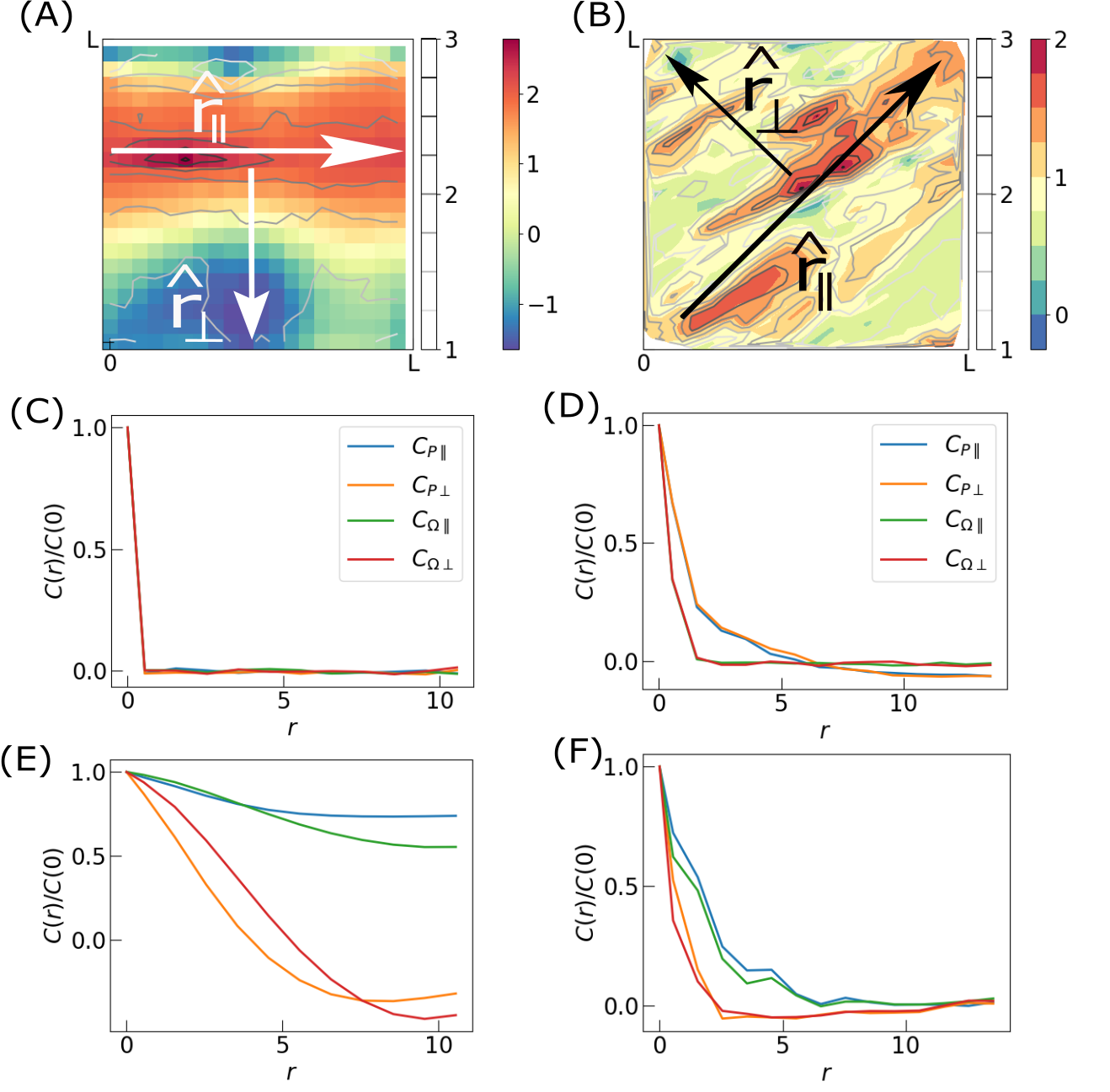}
\end{center}
\caption{
(A) Spatial map of $\Omega$ in the final step of a Monte Carlo simulation. Two mutually perpendicular directions are identified, $r_{\parallel}$ which is along the largest polarization domain, and $r_{\perp}$ which is perpendicular to it.
(B) Spatial map of $\Omega$ in the a steady-state configuration of the  vertex model simulation. Here $r_{\parallel}$ and $r_{\perp}$ are tilted by $45^{\circ}$ because a shear is applied to the tissue in the $xy$ direction.
(C) Correlations $C_P$ and $C_{\Omega}$ in parallel ($r_{\parallel}$) and perpendicular ($r_{\perp}$) directions in the initial state of Monte Carlo simulation. The initial state is completely random, hence the correlations die out just beyond $r=0$.
(D) Initial state of the vertex model simulations also show insignificant correlation.
(E) Long range correlation is observed in the final state of the Monte Carlo simulation. Correlation lengths in the parallel direction are longer than those in the perpendicular direction.
(F) Correlations in the final state of vertex model simulation also show larger correlation lengths in the parallel direction compared to those in the perpendicular direction.
The correlation plots have been produced by averaging over $50$ independent Monte Carlo runs (subfigures C and E), and $100$ independent Vertex model simulations (subfigures D and F).
}
    \label{fig:correlation}
\end{figure}

\begin{figure}[h!]
\begin{center}
\includegraphics[width=0.9\textwidth]{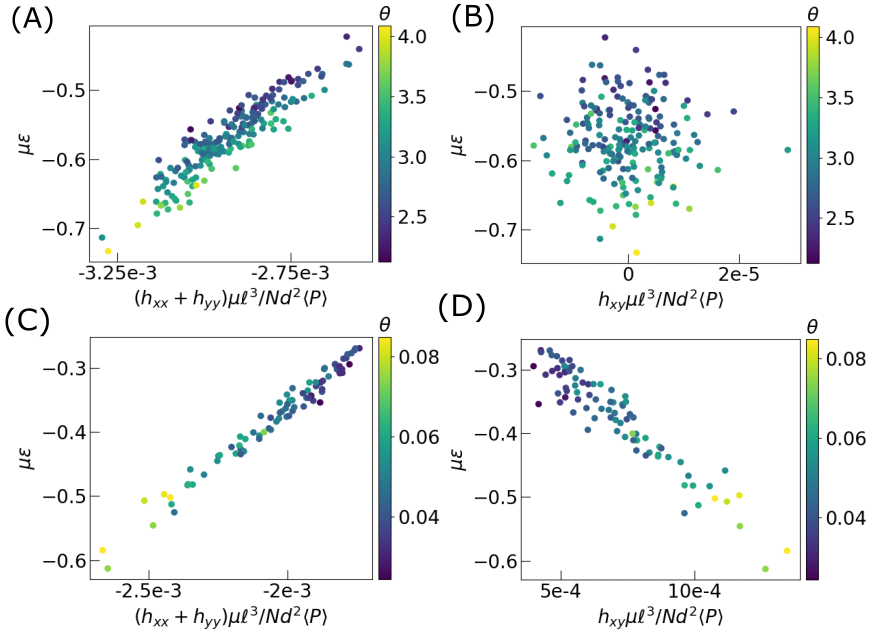}
\end{center}
\caption{ Energy versus average mean field. 
The $y$ axis represents dimensionless energy in units of $\mu$, and $x$ axis represents components of the dimensionless field tensor $h$. 
Each data point represents final time frame of a different experiment. 
The color of each point is the $\theta$ value of the corresponding polarization distribution. 
(A,B) Results from Monte Carlo, (C,D) Vertex model simulations. 
Energy is calculated using eq \ref{eq:interaction} for both models. 
In case of the vertex model, we have assumed the Poisson ratio of the tissue to be $0.5$. 
The imposed shear in the vertex model, in the $xy$ direction creates a strong effective field in that direction, which is absent in the Monte Carlo simulations without shear.}
    \label{fig:EvsH}
\end{figure}

We are specifically interested in understanding the differences/similarities between (a) correlations, and (b) energy landscapes that result from our mechanistic model and the vertex model. These comparisons are discussed in the context of Fig. \ref{fig:correlation}, and Fig. \ref{fig:EvsH}, respectively.  Our perspective is that both model cell shapes in tissues but ours is a purely mechanistic model, and comparison between the two can distinguish between mechanistic and geometrical influences on the spatial organization of cell shapes.
Both Monte Carlo and vertex model simulations show the common features of (i) formation of ordered domains, (ii) formation of domains of large polarization, and (iii) overlap of these two types of domains.

In order to quantify these observations we calculate spatial correlations of (a) the polarization magnitudes ($C_P$) and (b) the order parameter ($C_{\Omega}$) for the two  models (fig \ref{fig:correlation}).
In the initial states, the correlation of both $P$ and $\Omega$ die out very fast, indicating that there is no correlation in these variables to start with.
Since we observe formation of anisotropic domains with a long and a short axis in the final states, we calculate correlations in two independent directions - (i) parallel to the largest domain and (ii) perpedicular to it. 
Because of the externally imposed shear, the ``parallel''  direction is always at $45^{\circ}$ to the $x$ axis in case of the vertex model simulation.
In the case of Monte Carlo simulations, for each final state,  we align the largest domain along the $x$ axis, for convenience, and determine the average correlations over all final states.
In the final states of both the Monte Carlo simulations and the steady states of the vertex model simulations, $C_P$ and $C_\Omega$ decay simultaneously indicating a strong correlation between the two quantities. 
For the Monte Carlo runs we see a significant increase in correlation lengths of both quantities to almost $1/5$-th of the system size in perpendicular direction and system size in parallel direction.
This sharp contrast between longitudinal and transverse correlations is also evident in  Fig. (3c) depicting nematic order in \cite{armengol-collado_epithelia_2022}.  There it was remarked that these ``chain-like'' correlations are reminiscent of force-chains in granular media.  We want to emphasize that the origin of this type of correlation in the polarization that we observed is the same as that observed in stresses in granular media (\cite{Jishnu_PRL}).  In Fig. \ref{fig:correlation}(A,B)  the heatmap of $\Omega$ is analogous to a heatmap of  grain-level stresses shown in Fig.3 of \cite{Jishnu_PRL}, since $\Omega$ is a measure of $P_{xx}$.  The origin of  the longer-ranged correlations  of the dipole tensor $P_{\alpha \beta}$ in our model tissues and stress components in jammed granular solids is the imposition of the local constraint of force balance on each cell~\cite{bischofs_elastic_2004} and grain~\cite{Jishnu_PRL,Jishnu_PRE}. As discussed at length in the context of granular solids, this constraint leads to a Gauss's law type constraint in the continuum elasticity theory\cite{Jishnu_PRL,Jishnu_PRE}, leading to a pinch-point singularity in stress-stress correlations that implies much longer-ranged correlations in the longitudinal direction.
The Greens function in eqn. \ref{eq:interaction} encapsulates the force-balance constraint, and is directly responsible for the observed difference between longitudinal and transverse correlations. The visual appearance of ``force-chain'' like structures is a reflection of these correlations.

As  in the Monte Carlo simulations, the correlations observed in the steady-state of the vertex model (fig \ref{fig:correlation} (F)) are stronger in the parallel direction than those in the perpendicular direction.  
But the correlation lengths are systematically smaller in the vertex model.
We believe that this is due to a difference in cell dynamics in the two models. Cell dynamics in the vertex model simulation is subject to geometric constraints which prevent "long-range exchanges" of cell aspect ratios (polarization magnitues), allowed in Model 1 of the mechanistic model. These Monte Carlo moves in Model 1 facilitate formation of large ordered domains of polarization magnitudes, spanning the system size. 
As observed in fig \ref{fig:correlation} (B), the steady states of the vertex model are instead characterized by multiple domains of large polarization magnitudes.
This reduces the correlation length in the vertex model simulations compared to those in the Monte Carlo simulations. 
If instead of Model 1, we analyze the results of Model 2, where 
we spatially quench all the force dipoles in the Monte Carlo simulation, we observe several small domains, and much shorter correlations lengths (See Fig S1 in Supplementary Material).
We expect the dynamics in a physical tissue to lie somewhere in between the two extreme scenarios represented by Model 1 and Model 2, because there are additional constraints on how cells can migrate or change their geometrical shape. It is more likely that there is a characteristic length scale over which cells  can exchange polarization, which will define a domain size. 
\\

Lastly, we analyze the energy landscape explored at low temperatures by looking at the energies of the final states in the simulations.
In the Monte Carlo simulations, independent runs achieve different values of the final energy.
We observe, however, that if the simulation is run with a fixed set of $P$ but with different spatial realizations in the initial state,  they attain final states with  energy values that are virtually indistinguishable from each other.
This observation indicates  that the energy landscape at low temperatures is not sensitive to the initial configurations but is controlled by the particular realization of $P$, which are drawn from a $k-\Gamma$ distribution.
To quantify this feature, we plot the energy as a function of the average field $h_{\alpha \beta}$ (eqn \ref{eq:nematic}) in the final states.
Each point in fig \ref{fig:EvsH} represents one configuration, with its $y$ coordinate indicating total energy and $x$ coordinate indicating value of average field $h_{\alpha \gamma}=\frac{1}{N}\sum_i \sum_{i<j} G_{\alpha \beta, \gamma \beta} ({\bf r}_i-{\bf r}_j) P({\bf r}_j )$. 
We see a strong correlation between energy and the average local field $h_{xx}+h_{yy}$ in both Monte Carlo and vertex model simulations.  As seen from Eq.\ref{eq:nematic}, the average local field is controlled by the total polarization, $\sum_i P_i$.  We have checked that replacing average field $h_{\alpha \beta}$ by the total polarization produces plots that capture the same correlation as seen in Fig. \ref{fig:EvsH}.
The vertex model runs also show a strong correlation between final energy and the transverse field $h_{xy}$ because of the shearing in that direction, but this trend is absent in the Monte Carlo simulations, where the values of $h_{xy}$ are also much smaller because of the isotropy of the system.

\subsection{Energy Landscape}

\begin{figure}[h!]
\begin{center}
\includegraphics[width=\textwidth]{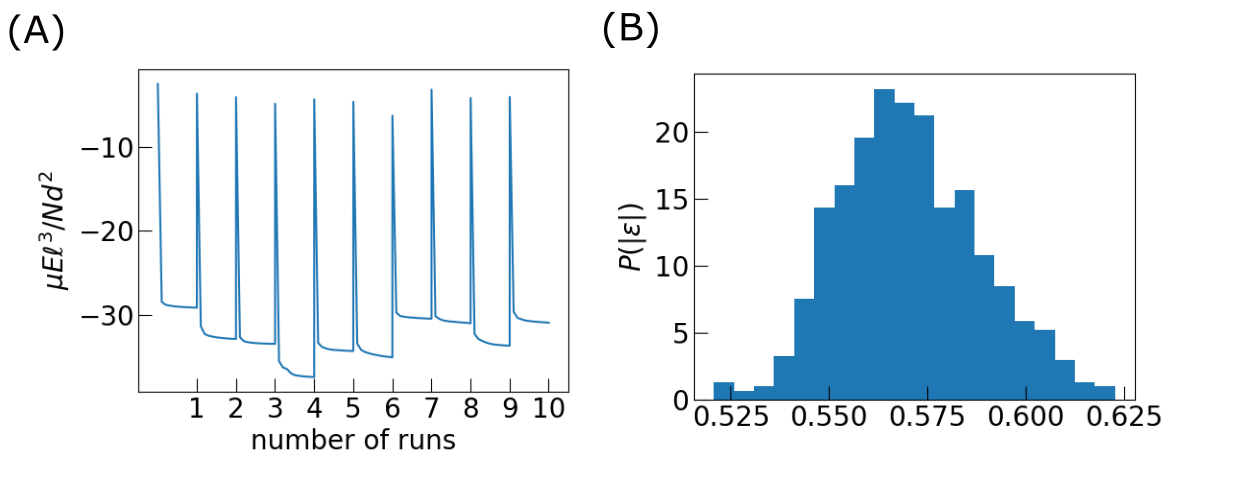}
\end{center}
\caption{ (A) In this figure we plot the time dependence of energy of the system for 10 independent MC runs. 
Each run is characterised by an effective value of $k$ and $\theta$. 
Each system reaches a different steady state energy depending on its specific $k, \theta$ value. 
(B) Distribution of normalized energies $\varepsilon$ obtained from the last configuration of each Monte Carlo run.}
    \label{fig:landscape}
\end{figure}

The other aspect of the model that we explore in this study is the energy landscape explored in Model 1.
Fig \ref{fig:landscape}(A) shows the time dependence of energy for ten independent Monte Carlo runs.
Each of our simulations is initiated with a set of polarization magnitudes drawn from a $k-\Gamma$ distribution which is then kept fixed throughout that run. 
The initial configuration is completely random and hence is a high energy state. 
As the simulation progresses, the system approaches a low energy configuration.
It is evident from the plot that each system approaches a different value of the final energy. 
The energy of the final state depends on the average value of polarization magnitudes of the dipoles. We performed three different sets of simulations fixing the value of $k$ to be $2.5$ but with $\theta=1.0, 2.0, 3.0$.
Since the systems are of finite size, each set of polarization magnitudes drawn from, nominally, the same $k-\Gamma$ distribution is characterized by $k, \theta$ values that are slightly different, leading to different values of the final energy.

If we normalize the final energies from all three sets of runs by the square of the average polarization of the respective run, then they fall in a pretty narrow range of values as shown by the distribution in fig \ref{fig:landscape}(B).

\begin{figure}[h!]
\begin{center}
\includegraphics[width=\textwidth]{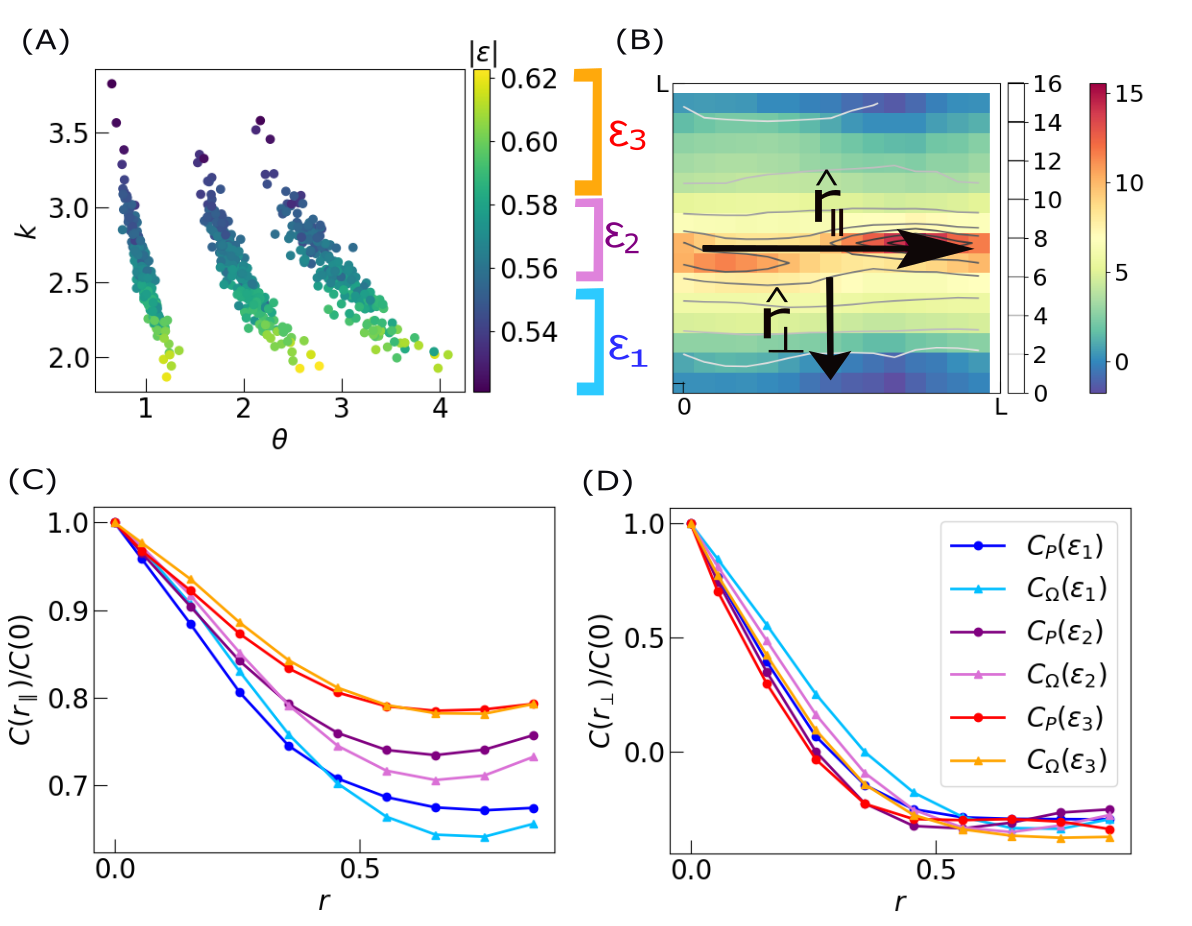}
\end{center}
\caption{ (A) Scatter plot of absolute values of final energy of each run against the respective $k, \theta$ values obtained from the particular distribution of P. 
The three groups of data correspond to simulations run at three sets of parameters $k=2.5$ and $\theta=1.0, 2.0, 3.0$. 
The actual values of $k, \theta$ in these finite size systems vary about these mean values. 
The scaled energy values in all three groups of points have the same range. 
We divide the data by the values of final scaled energy into three sections $\varepsilon_1, \varepsilon_2, \varepsilon_3$, such that each section has equal number of data points. 
(B) Schematic diagram showing $\hat{r}_{\parallel}$ and $\hat{r}_{\perp}$ directions with respect to the domain of largest polarization. 
(C) Correlation functions $C_P$ ad $C_{\Omega}$ in the parallel direction averaged over points belonging to the energy sectors $\varepsilon_1 < \varepsilon_2 < \varepsilon_3$. Both $p$ and $\Omega$ have the longest correlation length corresponding to $\varepsilon_3$, and shortest correlation length corresponding to $\varepsilon_1$. 
(D) Correlation functions calculated in the same way in the perpendicular direction. There is no significant effect of energy on the correlation length in this direction. }
    \label{fig:correlations}
\end{figure}

{
In fig \ref{fig:correlations}(A), we plot the normalized final energies of each system against their respective $k, \theta$ values.
In all three sets of data, the range of the \textit{normalized} energy is approximately the same.
We divide the data points into three sections by their energy values and plot correlations in two mutually perpendicular directions $\hat{r}_{\parallel}$ and $\hat{r}_{\perp}$.
$\hat{r}_{\parallel}$ is defined as the direction in which the largest polarization domain is aligned and correlation $C_{\parallel}$ is calculated by taking pairs of points lying on thin strips along this direction.
Similarly for $C_{\perp}$ we take pairs of points on thin strips along the perpendicular direction.
We observe that both $P$ and $\Omega$ have longer-ranged correlations in the parallel direction, as compared to the perpendicular one.
In addition, as we go to higher values of absolute energies (lower in the energy landscape) the systems become more strongly correlated in the parallel direction whereas energy has no significant effect on the correlations in the perpendicular direction. Unsurprisingly, the conclusion is that lower energy states are more ordered. }

\section{Conclusion}

Using a combination of numerical simulations and theoretical analysis, our study investigates the interplay between the spatial distributions of cell aspect ratios and orientational (nematic) order, as well as the complexity of the resulting energy landscape.

Our main finding reveals a strong correlation between nematic order and the magnitude of cell polarizations (aspect ratios), which are known to follow $k-\Gamma$ distributions that vary across different tissue types. Therefore, it is critical to analyze the spatial organization of both cell aspect ratios and orientations in order to fully understand tissue morphogenesis and development.

These results have significant implications for the field of tissue engineering and regenerative medicine, as they highlight the importance of understanding how cell shape and orientation impact tissue function and organization. By better understanding the complex interplay between these factors, we can develop more effective strategies for engineering functional tissues in vitro and for promoting tissue regeneration {\it in vivo}.


The mechanistic model used in the study has not been extended to explore the effects of external stresses on tissue shape changes. This means that the model does not take into account the influence of external factors that can affect tissue shape and organization.
In reality, the shape changes that occur in biological tissues are often the result of complex interactions between internal cellular processes and external mechanical and biochemical factors. For example, during embryonic development, the shape and positioning of organs are influenced by mechanical forces generated by the surrounding tissues and organs, as well as by the biochemical signals that regulate cell behavior.

Therefore, it is important to extend the mechanistic model used in the study to include the effects of external stresses on tissue shape changes. This would require incorporating additional parameters and variables into the model, such as the magnitude and direction of external forces, the stiffness and elasticity of the surrounding tissues, and the presence of biochemical signals that modulate cell behavior.
Such an extended model would provide a more realistic representation of tissue shape changes and could help to uncover the underlying mechanisms that regulate these processes. This could ultimately lead to a better understanding of how tissues develop, grow, and repair, as well as how they respond to various physiological and pathological conditions.

\section*{Conflict of Interest Statement}

The authors declare that the research was conducted in the absence of any commercial or financial relationships that could be construed as a potential conflict of interest.

\section*{Author Contributions}
  
KM and BC conceptualized the project. KM, BC and DB designed the study.  KM designed the simulation protocol.  KM and RR performed simulations.  KM analyzed data from Monte Carlo simulations and the Vertex model.  KM, BC and DB wrote different sections of the manuscript.  All authors contributed to manuscript revision, and have read and approved the submitted version. 

\section*{Funding}
KM and BC acknowledge financial support from Brandeis Bioinspired MRSEC through grant no : NSF-DMR 2011846. DB was supported in part by NSF DMR-2046683 and the Alfred P. Sloan Foundation.

\section*{Acknowledgments}
KM and BC acknowlege many useful discussions with Micheal D'Eon. The authors would like to thank Anh Nguyen and Junxiang Huang for providing vertex model simulation data. KM thanks Christopher Amey and Minu Varghese for useful discussions.

\bibliographystyle{Frontiers-Harvard}
\bibliography{frontiers_tissue_paper.bib}

\end{document}